# A spin-wave logic gate based on a width-modulated dynamic magnonic crystal

A.A. Nikitin[1,2,3*], A.B. Ustinov[1,3], A.A. Semenov[1], A.V. Chumak[2], A.A. Serga[2], V.I. Vasyuchka[2], E. Lähderanta[4], B.A. Kalinikos[1], and B. Hillebrands[2]

[1] *Department of Physical Electronics and Technology, St. Petersburg Electrotechnical University, St. Petersburg, 197376 Russia*

[2] *Fachbereich Physik and Landesforschungszentrum OPTIMAS, Technische Universität Kaiserslautern, Kaiserslautern, 67663 Germany*

[3] *Department of Mathematics and Physics, Lappeenranta University of Technology, Lappeenranta, 53850 Finland*

An electric current controlled spin-wave logic gate based on a width-modulated dynamic magnonic crystal is realized. The device utilizes a spin-wave waveguide fabricated from a single-crystal Yttrium Iron Garnet film and two conducting wires attached to the film surface. Application of electric currents to the wires provides a means for dynamic control of the effective geometry of the waveguide and results in a suppression of the magnonic band gap. The performance of the magnonic crystal as an AND logic gate is demonstrated.

* and.a.nikitin@gmail.com

In recent years artificially patterned magnetic media, magnonic crystals (MCs), have commanded increased interest (see reviews [1-4] and literature therein). One of the important features of MCs is the presence of band gaps in the spin-wave spectrum, i.e. frequency bands in which the propagation of spin waves is forbidden [5-7]. The dispersion management introduced by an artificial periodicity of the magnetic film allows for the observation of a variety of different linear and nonlinear spin-wave phenomena. Among them are the formation of band gaps [8-11] and the generation of gap solitons [12, 13]. Magnonic crystals are also promising for practical applications, particularly for phase shifters [14] and generators [15], and they have also been suggested for use as temperature or magnetic field sensors [16, 17].

A very fascinating type of MCs is the dynamic magnonic crystal (DMC) [18-20]. This is a wave guiding structure with rapidly switchable periodic properties. Up to now, it has been realized in the form of a YIG film with a meander type conductor placed on its surface. An electric current in the conductor produces a periodic spatial modulation of the bias magnetic field [18]. Due to this property, the dynamic magnonic crystal allows for unique signal processing functions, such as all-linear time reversal, frequency conversion [19], and signal storage [20].

In this Letter, we report on the realization of a width-modulated dynamic magnonic crystal (WMDMC). In the designed structure, in contrast to the aforementioned DMC, electric currents control the effective geometry of the magnetic film waveguide for spin waves propagating therein. Moreover, we demonstrate the application of the WMDMC as a spin-wave logic gate.

It should be noted that so far spin-wave logic gates have only been realized using a spin-wave interferometer geometry [21-26]. For example, elements providing controllable phase shifts for spin waves propagating in the arms of a Mach-Zehnder interferometer were included in each arm. The logic operations were based on changing the phase of the spin-wave by $\pi$ with an external current signal coded as a logical '1'. The logic functionality can also be realized via control of the spin-wave amplitude in the interferometer arms [23, 26]. The logic gate presented here does not utilize any interferometer circuitry; instead, it consists of one WMDMC, which controls the spin-wave amplitude. This function is realized through control of the sensitivity of the propagating spin wave to the edge modulation of the magnetic-film waveguide structure.

A schematic view of the WMDMC is shown in Fig. 1. The device was fabricated using an epitaxial Yttrium Iron Garnet (YIG) film. For the experimental investigation, the YIG-film waveguide was cut from a high-quality single-crystal YIG film of 8.5 μm thickness grown on 500 μm thick Gadolinium Gallium Garnet (GGG) substrate by liquid-phase epitaxy. The film has pinned surface spins that lead to standing spin-wave resonances and appearance of additional dips in the amplitude-frequency characteristic [27]. Periodic sinusoidal width modulation of the YIG-film waveguide was made by chemical etching of the film. The spatial modulation period $T_m$ and the peak-to-peak amplitude $2A_m$ (see Fig. 1) were 400 μm and 200 μm, respectively (see Fig. 1). The length of the modulated area was 4 mm, while the width of the unmodulated section of the YIG-film waveguide was 1.5 mm. A spatially uniform bias magnetic field was applied across the YIG waveguide in order to provide the conditions for excitation of surface magnetostatic spin waves [28]. These waves were exited and detected by the microwave strip-line antennas placed at equal distances from both ends of the modulated YIG-film area 10 mm away from each other. Two gold wires of 50 µm in diameter were placed on the YIG-film surface along the modulated edges, as shown in Fig. 1. These wires were used for the dynamic control of the magnonic crystal band gap by supplying the electric currents.

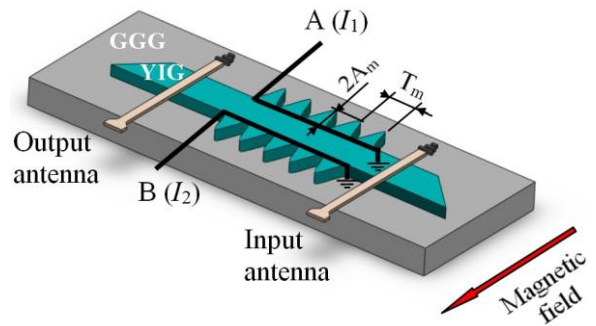

Fig. 1. Sketch of the width-modulated dynamic magnonic crystal. The ports A and B represent the logical inputs. The currents $I_1$ and $I_2$ supplied to the ports correspond to a logical '1' if their values equal $I_{MBG}$. Zero currents correspond to a logical '0'. The output spin-wave signal represents the logical output.

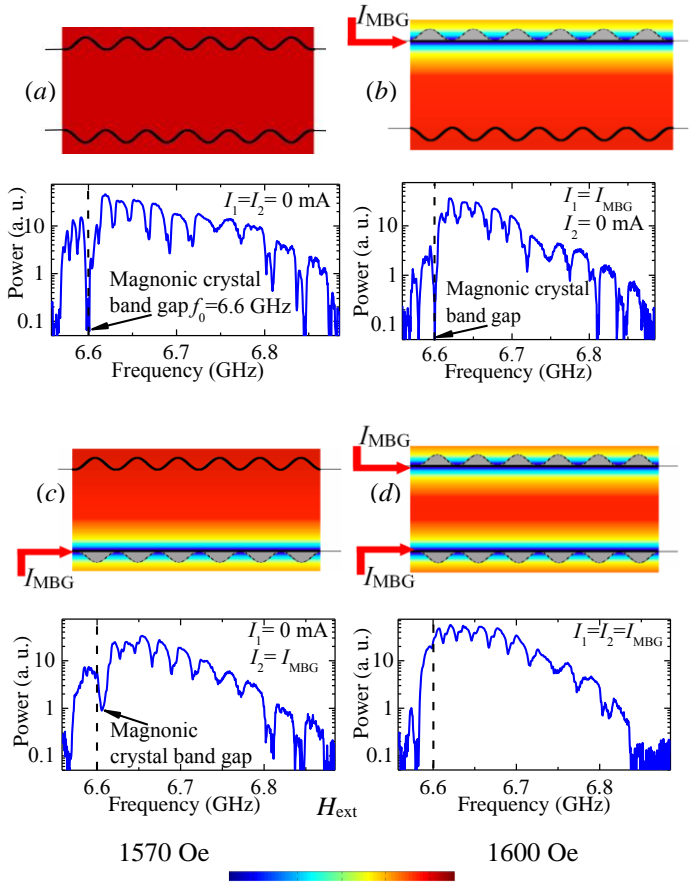

Fig. 2. Distributions of the bias magnetic field (top) and transmitted spin-wave power versus frequency for the width-modulated dynamic magnonic crystal (bottom) for the cases when the electric current is not applied to the wires (a); the current $I_{MBG}$ is applied to only one of the wires (b,c); and the current $I_{MBG}$ is applied to both wires (d). The value of the current $I_{MBG}$ = 1500 mA.

Let us consider qualitatively the formation of a band gap in the width-modulated magnonic crystal. Generally, a wave vector of traveling spin waves exited by the input antenna in the YIG film has two components, which are in-plane. The first component is longitudinal along the direction from the input to the output antenna while the second component is transverse. As it is physically clear, the value of the transverse component is dependent on the width of the waveguide. This effect leads to the periodic change in the wave guiding properties of the width-modulated YIG film and provides the magnonic band gap (MBG) in the amplitude vs. frequency characteristic of the investigated MC. The number of MBGs is defined by the type of the width modulation. For example, in the case of rectangular modulation there are several MBGs. As soon as the width modulation is sinusoidal, as is the case in the investigated structure (Fig. 1), there is only one MBG in the power vs. frequency characteristic (see Fig. 2(a)) [29].

The physical mechanism underlying the control of the magnonic crystal band gap by the electric currents applied to the wires can be understood as follows. Positive currents $I_1$ and $I_2$ applied to both wires create an additional negative Oersted field with respect to the applied static bias field. Therefore, the spatial distribution of the resulting magnetic field has two minima at the borders of the width modulation. The distributions of the bias magnetic field are shown in Fig. 2 as color maps. Provided the depth of the minima is sufficiently large, they screen the YIG-film modulated edges due to a decrease in the magnetic field inside the YIG waveguide. In other words, the existence of these minima cause the width modulation of the waveguide to become "invisible" to spin waves. This effect leads to the disappearance of the band gap at $f_0$ = 6.6 GHz, as is shown in Fig. 2(d).

We will now present the dynamic properties of the device structure realized upon application of the controlling dc electric currents. The experiments were carried out in the following manner. A microwave signal applied to the input antenna excited a surface magnetostatic spin wave. After passing through the magnonic crystal the spin wave was detected by the output antenna (Fig. 1). The power of the received spin waves was measured as a function of their carrier frequency. A typical dependence measured for $H$ = 1600 Oe with no current in the edge wires is shown in Fig. 2(a). As can be seen in the graph, the center frequency of the band gap was $f_0$ = 6.6 GHz.

The dependence of the power of the output signal vs. frequency obtained under two equal controlling currents $I_{MBG}$ of 1500 mA is shown in Fig. 2(d). Note that the data clearly demonstrate (in comparison to Fig. 2(a)) a suppression of the band gap at 6.6 GHz. If the dc current was applied to one of the wires then only one of the modulated edges was deactivated. In this case, the input spin wave was still scattered on its path through the waveguide. In this situation, the band gap appears as shown in Figs 2(b, c). The difference between Fig. 2(b) and Fig. 2(c) could be explained by a possible asymmetry in the distribution of the spin-wave amplitude along the cross-section of the YIG-film waveguide. In this case, there is a difference in the sensitivity of the propagating wave to the boundary conditions on the different waveguide edges.

Dependences of the spin-wave transmission as a function of the currents $I_1$ and $I_2$ measured at the frequency $f_0$ = 6.6 GHz is shown in Fig. 3. As is visible, an increase in both currents leads to a substantial increase in the spin-wave transmission due to a reduction in the influence of the modulated edges. At the same time, there is a relatively weak influence with only one dc current applied.

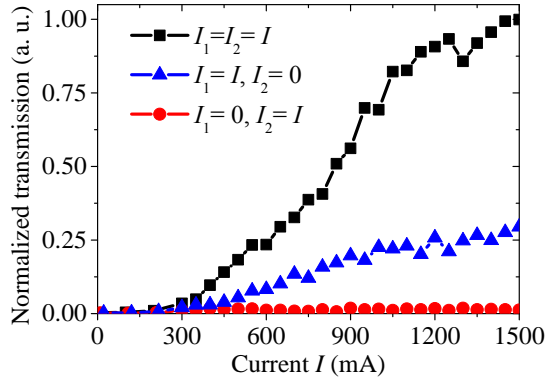

Fig. 3 Dependence of the spin wave transmission on the value of the dc current at frequency $f_0$=6.6 GHz. Squares: both currents $I_1$ and $I_2$ are changed; triangles: the current $I_1$ is changed and $I_2$=0; circles: $I_1$=0 and $I_2$ is changed.

A possible application of the WMDMC for realization of an AND logic gate will now be explored. The operating principle of the AND logic gate is based on controlling the bias magnetic field distribution along the sinusoidal borders of the YIG-film waveguide through a change in the electric currents $I_1$ and $I_2$. The logic '0' is represented by $I = 0$ mA and a logic '1' − by $I = I_{MBG} = 1500$ mA, which is enough to "switch off" the width modulation on one side of the WMDMC. The microwave pulses at the input antenna represent clock pulses. The microwave signal at the output antenna serves as the logic signal output. The operational frequency corresponds to the central frequency of the band gap, for example $f_0 = 6.6$ GHz. Therefore, the presence of the band gap (i.e., low power level at the output) represents a logic '0' and the absence of the band gap (i.e., high power at the output, $P_{out} = P_1$) represents a logic '1'.

Typical oscilloscope traces of the output signal demonstrating the performance of the logic gate are shown in Fig. 4. It is observed that the logic '0' appears at the output port in three situations: for two logic '0s' applied to the input ports A and B (Fig. 4(*a*)) and for the combinations of the logic '0' and the logic '1' applied to the inputs (Fig. 4(*b*, *c*)). In the case where a logic '1' is applied simultaneously to both inputs, the signal at the output port corresponds to a logic '1'.

In conclusion, width-modulated magnonic crystals may find a variety of applications. For example, they can be employed for the development of dynamic magnonic crystals and spin-wave logic gates, as demonstrated in this paper. Another example could be microwave notch filters with a stop-band frequency corresponding to the magnonic band gap. Advantages of the dynamic magnonic crystal presented above in comparison to the ones previously developed [18-20] are the potentially fast performance and the possibility for miniaturization. Indeed, the current control is provided here by short conductors, which have an inductance much less than the meander type wire reported in [18]. The use of nanostructured width-modulated waveguides made of Permalloy films, as in Ref. [9], will allow for a significant reduction of the size of the dynamic magnonic crystal and the logic circuit presented here.

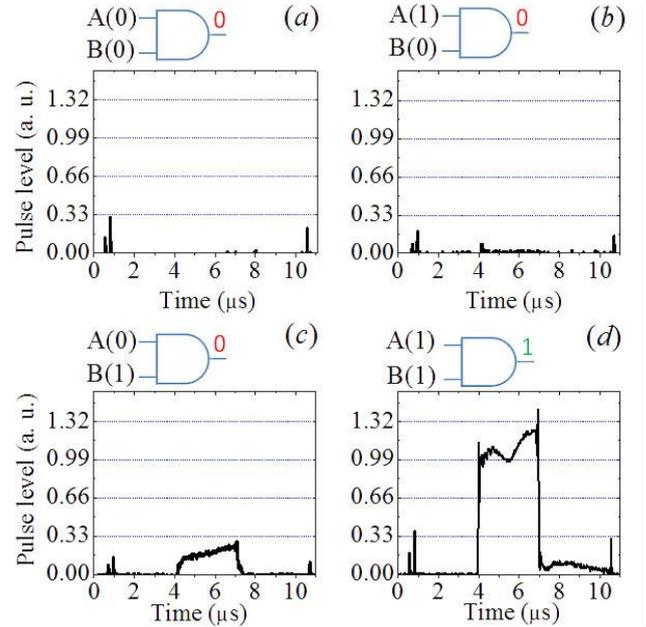

Fig. 4 Waveforms of the output signal of the AND logic gate. Microwave signal frequency $f_0$=6.6 GHz. Currents $I_1$ and $I_2$ serve as logic inputs; logic '0' corresponds to zero current, logic '1'- to $I_{MBG}$=1500 mA. High level of the microwave output signal corresponds to a logic '1', and low level to a logic '0'.

The work was supported in part by the Russian Science Foundation (Grant 14-12-01296), the Ministry of Education and Science of Russian Federation, the Academy of Finland, EU-FET grant InSpin 612759, and by the Deutsche Forschungsgemeinschaft.